
\newcount\mgnf\newcount\tipi\newcount\tipoformule
\newcount\driver
\driver=1
\mgnf=0 
\tipi=2 
\tipoformule=0 
\ifnum\mgnf=0 \magnification=\magstep0\hoffset=0.cm
   \hsize=15.5truecm\vsize=22.truecm \parindent=10.pt\fi
\ifnum\mgnf=1 \magnification=\magstep1\hoffset=0.truecm
   \voffset=-0.5truecm\hsize=16.5truecm\vsize=24.truecm
   \baselineskip=14pt plus0.1pt minus0.1pt \parindent=12pt
   \lineskip=4pt\lineskiplimit=0.1pt \parskip=0.1pt plus1pt\fi
%
%
\let\a=\alpha  \let\g=\gamma \let\d=\delta
\let\e=\varepsilon
\let\z=\zeta \let\h=\eta   \let\l=\lambda
 \let\n=\nu \let\x=\xi  \let\r=\rho
\let\s=\sigma \let\t=\tau  
\let\ps=\psi \let\o=\omega  
  \let\Th=\Theta \let\L=\Lambda
  \let\F=\Phi


{\count255=\time\divide\count255 by 60 \xdef\oramin{\number\count255}
        \multiply\count255 by-60\advance\count255 by\time
        \xdef\oramin{\oramin:\ifnum\count255<10 0\fi\the\count255}}

\def\ora{\oramin }

\def\data{\number\day/\ifcase\month\or gennaio \or febbraio \or marzo
\or aprile \or maggio \or giugno \or luglio \or agosto \or settembre
\or ottobre \or novembre \or dicembre \fi/\number\year;\ \ora}

\setbox200\hbox{$\scriptscriptstyle \data $}

\newcount\pgn \pgn=1
\def\foglio{\number\numsec:\number\pgn
\global\advance\pgn by 1}  \def\foglioa{A\number\numsec:\number\pgn
\global\advance\pgn by 1}

\newskip\ttglue
\def\TIPI{
\font\ottorm=cmr8   \font\ottoi=cmmi8
\font\ottosy=cmsy8  \font\ottobf=cmbx8
\font\ottott=cmtt8  
\font\ottoit=cmti8
\def \ottopunti{\def\rm{\fam0\ottorm}
\textfont0=\ottorm  \textfont1=\ottoi
\textfont2=\ottosy  \textfont3=\ottoit
\textfont4=\ottott
\textfont\itfam=\ottoit  \def\it{\fam\itfam\ottoit}%
\textfont\ttfam=\ottott  \def\tt{\fam\ttfam\ottott}%
\textfont\bffam=\ottobf
\normalbaselineskip=9pt\normalbaselines\rm}
\let\nota=\ottopunti}
\def\TIPIO{
\font\setterm=amr7 
\font\settesy=amsy7 \font\settebf=ambx7 
\def \settepunti{\def\rm{\fam0\setterm}
\textfont0=\setterm   
\textfont2=\settesy   
\textfont\bffam=\settebf  \def\bf{\fam\bffam\settebf}
\normalbaselineskip=9pt\normalbaselines\rm
}\let\nota=\settepunti}

\def\TIPITOT{
\font\twelverm=cmr12
\font\twelvei=cmmi12
\font\twelvesy=cmsy10 scaled\magstep1
\font\twelveex=cmex10 scaled\magstep1
\font\twelveit=cmti12
\font\twelvett=cmtt12
\font\twelvebf=cmbx12
\font\twelvesl=cmsl12
\font\ninerm=cmr9
\font\ninesy=cmsy9
\font\eightrm=cmr8
\font\eighti=cmmi8
\font\eightsy=cmsy8
\font\eightbf=cmbx8
\font\eighttt=cmtt8
\font\eightsl=cmsl8
\font\eightit=cmti8
\font\sixrm=cmr6
\font\sixbf=cmbx6
\font\sixi=cmmi6
\font\sixsy=cmsy6
\font\twelvetruecmr=cmr10 scaled\magstep1
\font\twelvetruecmsy=cmsy10 scaled\magstep1
\font\tentruecmr=cmr10
\font\tentruecmsy=cmsy10
\font\eighttruecmr=cmr8
\font\eighttruecmsy=cmsy8
\font\seventruecmr=cmr7
\font\seventruecmsy=cmsy7
\font\sixtruecmr=cmr6
\font\sixtruecmsy=cmsy6
\font\fivetruecmr=cmr5
\font\fivetruecmsy=cmsy5
\textfont\truecmr=\tentruecmr
\scriptfont\truecmr=\seventruecmr
\scriptscriptfont\truecmr=\fivetruecmr
\textfont\truecmsy=\tentruecmsy
\scriptfont\truecmsy=\seventruecmsy
\scriptscriptfont\truecmr=\fivetruecmr
\scriptscriptfont\truecmsy=\fivetruecmsy
\def \eightpoint{\def\rm{\fam0\eightrm}
\textfont0=\eightrm \scriptfont0=\sixrm \scriptscriptfont0=\fiverm
\textfont1=\eighti \scriptfont1=\sixi   \scriptscriptfont1=\fivei
\textfont2=\eightsy \scriptfont2=\sixsy   \scriptscriptfont2=\fivesy
\textfont3=\tenex \scriptfont3=\tenex   \scriptscriptfont3=\tenex
\textfont\itfam=\eightit  \def\it{\fam\itfam\eightit}%
\textfont\slfam=\eightsl  \def\sl{\fam\slfam\eightsl}%
\textfont\ttfam=\eighttt  \def\tt{\fam\ttfam\eighttt}%
\textfont\bffam=\eightbf  \scriptfont\bffam=\sixbf
\scriptscriptfont\bffam=\fivebf  \def\bf{\fam\bffam\eightbf}%
\tt \ttglue=.5em plus.25em minus.15em
\setbox\strutbox=\hbox{\vrule height7pt depth2pt width0pt}%
\normalbaselineskip=9pt
\let\sc=\sixrm  \let\big=\eightbig  \normalbaselines\rm
\textfont\truecmr=\eighttruecmr
\scriptfont\truecmr=\sixtruecmr
\scriptscriptfont\truecmr=\fivetruecmr
\textfont\truecmsy=\eighttruecmsy
\scriptfont\truecmsy=\sixtruecmsy
}\let\nota=\eightpoint}

\newfam\msbfam   
\newfam\truecmr  
\newfam\truecmsy 
\newskip\ttglue
\ifnum\tipi=0\TIPIO \else\ifnum\tipi=1 \TIPI\else \TIPITOT\fi\fi



\def\V#1{{\vec#1}}\let\dpr=\partial\let\ciao=\bye
\let\io=\infty

\def\media#1{\langle{#1}\rangle} \let\0=\noindent

\def\guida{\leaders\hbox to 1em{\hss.\hss}\hfill}
\def\tende#1{\vtop{\ialign{##\crcr\rightarrowfill\crcr
              \noalign{\kern-1pt\nointerlineskip}
              \hglue3.pt${\scriptstyle #1}$\hglue3.pt\crcr}}}
\def\otto{{\kern-1.truept\leftarrow\kern-5.truept\to\kern-1.truept}}


\def\ie{\hbox{\sl i.e.\ }}
\def\eg{\hbox{\sl e.g.\ }}
\def\qed{\raise1pt \hbox{\vrule height5pt width5pt depth0pt} }

\def\fiat{{}}




\def\HH{{\cal H}}

\def\TT{{\cal T}}

\def\={{ \; \equiv \; }}


\let\dt=\displaystyle

\def\2{{1\over2}}

\def\igb{
\mathop{\raise4.pt\hbox{\vrule height0.2pt depth0.2pt width6.pt}
\kern0.3pt\kern-9pt\int}}

\def\fra#1#2{{#1\over#2}}
\let\\=\noindent

\def\*{\vskip0.5truecm}

\font\titolo=cmbx12

\def\nn{{\vec \n}}
\def\pps{{\vec \psi}}
\def\oo{{\vec \o}}
\def\AA{{\vec A}}

\def\HH{{\vec H}}
\def\hh{{\vec h}}
\def\aa{{\vec \a}}

\def\aaa{{\vec a}}

\newdimen\xshift\newdimen\xwidth\newdimen\yshift\def\eqfig#1#2#3#4#5{\par
\xwidth=#1\xshift=\hsize\advance\xshift by-\xwidth\divide\xshift by 2
\yshift=#2\divide\yshift by 2\line{\hglue\xshift\vbox to #2{\vfil
\ifnum\driver=0 #3
\special{ps: plotfile #4.ps}
\fi\ifnum\driver=1
#3\includegraphics{#4.ps}\fi\ifnum\driver=2 #3\special{\ifnum\mgnf=0#4.ps
1. 1. scale\fi\ifnum\mgnf=1 #4.ps 1.2 1.2 scale\fi}\fi}\hfill
\raise\yshift\hbox{#5}}}
\def\ins#1#2#3{\vbox to0pt{\kern-#2\hbox{\kern#1 #3}\vss}\nointerlineskip}
\def\insertplot#1#2#3#4{
    \par \xwidth=#1 \xshift=\hsize \advance\xshift
     by-\xwidth \divide\xshift by 2 \vbox{
  \line{} \hbox{ \hskip\xshift  \vbox to #2{\vfil
 \ifnum\driver=0 #3  
                 \special{ps: plotfile #4.ps} 
 \fi
 \ifnum\driver=1  #3    \includegraphics{#4.ps}       \fi
 \ifnum\driver=2  #3   \ifnum\mgnf=0
                       \special{#4.ps 1. 1. scale}\fi
                       \ifnum\mgnf=1
                       \special{#4.ps 1.2 1.2 scale}\fi\fi
 \ifnum\driver=5  #3   \fi}
\hfil}}}

\def\figini#1{
\def\8{\write13}
\catcode`\%=12\catcode`\{=12\catcode`\}=12
\catcode`\<=1\catcode`\>=2
\openout13=#1.ps}

\def\figfin{
\closeout13
\catcode`\%=14\catcode`\{=1
\catcode`\}=2\catcode`\<=12\catcode`\>=12
}

\global\newcount\numsec\global\newcount\numfor
\global\newcount\numapp\global\newcount\numcap
\global\newcount\numfig\global\newcount\numpag
\global\newcount\numnf

\def\SIA #1,#2,#3 {\senondefinito{#1#2}%
\expandafter\xdef\csname #1#2\endcsname{#3}\else
\write16{???? ma #1,#2 e' gia' stato definito !!!!}\fi}

\def\FU(#1)#2{\SIA fu,#1,#2 }

\def\etichetta(#1){(\veroparagrafo.\veraformula)%
\SIA e,#1,(\veroparagrafo.\veraformula) %
\global\advance\numfor by 1%
\write16{ EQ #1 ==>\equ(#1)  }}
\def\etichettaa(#1){(A\veraappendice.\veraformula)
\SIA e,#1,(A\veraappendice.\veraformula)
\global\advance\numfor by 1
\write16{ EQ #1 ==>\equ(#1) }}
\def\getichetta(#1){Fig.\verafigura
\SIA g,#1,{\verafigura}
\global\advance\numfig by 1
\write16{ Fig. #1 ==>\graf(#1) }}
\def\retichetta(#1){\numpag=\pgn\SIA r,#1,{\verapagina}
\write16{\rif(#1) ha simbolo  #1  }}
\def\etichettan(#1){(n\verocapitolo.\veranformula)
\SIA e,#1,(n\verocapitolo.\veranformula)
\global\advance\numnf by 1
\write16{\equ(#1) <= #1  }}

\newdimen\gwidth
\gdef\profonditastruttura{\dp\strutbox}
\def\senondefinito#1{\expandafter\ifx\csname#1\endcsname\relax}
\def\BOZZA{
\def\alato(##1){
 {\vtop to\profonditastruttura{\baselineskip
\profonditastruttura\vss
\rlap{\kern-\hsize\kern-1.2truecm{$\scriptstyle##1$}}}}}
\def\galato(##1){\gwidth=\hsize\divide\gwidth by 2
 {\vtop to\profonditastruttura{\baselineskip
\profonditastruttura\vss
\rlap{\kern-\gwidth\kern-1.2truecm{$\scriptstyle##1$}}}}}
\def\verapagina{
{\romannumeral\number\numcap}.\number\numsec.\number\numpag}}

\def\alato(#1){}
\def\galato(#1){}
\def\veroparagrafo{\number\numsec}\def\veraformula{\number\numfor}
\def\veraappendice{\number\numapp}
\def\verapagina{\number\pageno}\def\veranformula{\number\numnf}
\def\verafigura{{\romannumeral\number\numcap}.\number\numfig}
\def\verocapitolo{\number\numcap}\def\veranformula{\number\numnf}
\def\Eqn(#1){\eqno{\etichettan(#1)\alato(#1)}}
\def\eqn(#1){\etichettan(#1)\alato(#1)}

\def\Eq(#1){\eqno{\etichetta(#1)\alato(#1)}}
\def\eq(#1){\etichetta(#1)\alato(#1)}
\def\Eqa(#1){\eqno{\etichettaa(#1)\alato(#1)}}
\def\eqa(#1){\etichettaa(#1)\alato(#1)}
\def\dgraf(#1){\getichetta(#1)\galato(#1)}
\def\drif(#1){\retichetta(#1)}

\def\eqv(#1){\senondefinito{fu#1}$\clubsuit$#1\else\csname fu#1\endcsname\fi}
\def\equ(#1){\senondefinito{e#1}\eqv(#1)\else\csname e#1\endcsname\fi}
\def\graf(#1){\senondefinito{g#1}\eqv(#1)\else\csname g#1\endcsname\fi}
\def\rif(#1){\senondefinito{r#1}\eqv(#1)\else\csname r#1\endcsname\fi}

\openin14=\jobname.aux\ifeof14\relax\else
\input\jobname.aux\closein14\fi


\fiat

\def\ot{\leftarrow}\def\HH{{\cal H}}\def\hh{{\sl h}}
\let\Eq=\eqno\def\eq{{}}\def\equ(#1){{Eq. (#1)}}
\vglue2.truecm

\\{\titolo Field theory and KAM tori}\footnote{${}^@$}{Archived in
{\it mp$\_$arc@math.utexas.edu}, \#95-151; in {\it chao-dyn@xyz.lanl.gov},
\# 9503???; last version at http://chimera.roma1.infn.it}
\vskip1truecm

\\{\bf G. Gallavotti,\footnote{${}^*$}{\nota
Dipartimento di Fisica, $I^a$ Universit\`a di Roma, P.le Moro 2,
00185 Roma, Italia; partially supported by Rutgers U. and CNR-GNFM.}
G. Gentile,\footnote{${}^\#$}{\nota IHES, 91440 Bures s/Yvette, France.},
V. Mastropietro\footnote{${}^\&$}{\nota
Dipartimento di Matematica, $II^a$ Universit\`a di Roma,
00133 Roma, Italia; supported from CNR-GNFM.}}

\* \\{\sl Abstract: The parametric equations of KAM tori for a $l$
degrees of freedom quasi integrable system, are shown to be one point
Schwinger functions of a suitable euclidean quantum field theory on the
$l$ dimensional torus. KAM theorem is equivalent to a ultraviolet
stability theorem.  A renormalization group treatment of the field
theory leads to a resummation of the formal pertubation series and to an
expansion in terms of $l^2$ new parameters forming a $l\times l$ matrix
$\s_\e$ (identified as a family of renormalization constants).  The
matrix $\s_\e$ is an analytic function of the coupling $\e$ at small
$\e$: the breakdown of the tori at large $\e$ is speculated to be
related to the crossing by $\s_\e$ of a ``critical" surface at a value
$\e=\e_c$ where the function $\s_\e$ is still finite. A mechanism for
the possible universality of the singularities of parametric equations
for the invariant tori, in their parameter dependence as well as in the
$\e_c-\e$ dependence, is proposed.}

\vskip1.5truecm

\\{\bf 1. Introduction}
\*\numsec=1\numfor=1

\\We consider $l$ rotators with inertia moments $J$, angular momenta
$\AA=(A_1,\ldots,A_l)\in{\bf R}^l$, and angular positions
$\aa=(\a_1,\ldots,\a_l)\in{\bf T}^l$. Their motion will be described by
the hamiltonian
$$ \eqalign{
H & = {1\over2}J^{-1}\AA\cdot\AA + \e \, f(\aa,\AA) \; , \qquad
\AA\in{\bf R}^l, \aa\in{\bf T}^l \; , \cr
f & = \sum_{|\nn|\le N} f_\nn(\AA) \, e^{i \nn\cdot\aa } \; ,
\qquad f_\nn(\AA) = f_{-\nn}(\AA) \; , \cr} \Eq(1.1) $$
with $f_\nn(\AA)$ a polynomial in $\AA$.\footnote{${}^1$}{\nota
Analyticity of $f$ in a domain $W(\AA_0,\r)=
\{\AA\in{\bf R}^l\,:\,|\AA-\AA_0|/|\AA_0|<\r\}$
would make the matter more complicate only slightly.}  Let
$\oo_0=J^{-1}\AA_0$ be a rotation vector, ``angular velocity vector",
verifying for $C_0, \t>0$ suitably chosen the {\sl diophantine property}
$$ C_0|\oo_0\cdot\nn|>|\nn|^{-\t} \, , \qquad
\V0 \neq \nn \in {\bf Z}^l \; . \Eq(1.2) $$

The KAM theorem states the existence of a one parameter family
$\e\to\TT_\e$ of tori with parametric equations
$$ \AA=\AA_0+\V H(\pps) \; , \qquad
\aa=\pps+\V h(\pps) \; , \quad \pps\in{\bf T}^l \; , \Eq(1.3) $$
where $\V H(\pps)$ and $\V h(\pps)$ are analytic functions
of $\e$, $\ps_j$, $j=1,\ldots,l$, divisible by $\e$,
defined for $|\e|,|\hbox{Im}\,\ps_j|$ small enough. Such tori
are uniquely determined by the requirements:
$$ \eqalign{
\hbox{(a) } & \quad
\pps \to \pps+\oo_0t \hbox{ solves the equations of motion} \; , \cr
\hbox{(b) } & \quad
H(\pps) \hbox{ is even in $\pps$} \; , \cr
\hbox{(c) } & \quad
\hh(\pps) \hbox{ is odd in $\pps$} \; , \cr} \Eq(1.4) $$

Consider the four (formally) gaussian vector fields $\V \F\=(\V
\HH^\s,\V \hh^\s)$, $\s=\pm$, defined on the torus ${\bf T}^l$, and
with propagators\footnote{${}^2$}{\nota \ie  linear functionals on
the space of complex felds on ${\bf T}^l$ such that the moments are
evaluated by using the Wick rule.}
$$\eqalign{
\langle \hh^{+}_{\pps,j}\hh^{-}_{\pps',j'} \rangle =& \d_{j,j'}
\sum_{\nn} {e^{i(\pps-\pps')\cdot\nn}
\over(i\oo_0\cdot\nn+\L^{-1})^2} \=
\d_{j,j'}\,S^2(\pps-\pps') \; , \cr
\media{\hh^{2+}_{\pps,j} \HH^{-}_{\pps',j'} } =&
\media{\HH^{+}_{\pps,j}\hh^{-}_{\pps',j'} }= \d_{j,j'}
\sum_{\nn}{e^{i(\pps-\pps')\cdot\nn}
\over(i\oo_0\cdot\nn+\L^{-1})}\=
\d_{j,j'}\,S^1(\pps-\pps') \; , \cr}\Eq(1.5) $$
where $\L$ is a {\it ultraviolet cut off}.\footnote{${}^3$}{\nota
Because $\oo_0\cdot\nn$, $\nn\neq\V0$, can become small
only for $|\nn|$ large.} The
other propagators are taken to be zero. The physical dimensions of
the field $\V \hh^{+}$, $\V \hh^{-}$, $\V \HH^{+}$, $\V \HH^{-}$ are
respectively $[1],[\o^{-2}],[\o],[\o^{-1}]$ in terms of the dimension
$[\o]$ of $\oo_0$. We sall also set $\V \F^{1\pm}\=\HH^\pm$ and $\V
\F^{2\pm}\=\hh^\pm$.

We denote by  $P(d\F)$ the formal functional integral with respect
to the above gaussian process, and consider the field theory with $\V\F$
as free field and {\it action}
$$ \eqalign{ V(\F) = & -\e \int_{{\bf T}^l} d\pps \,
J^{-1}\hh^{-}_\pps\cdot\dpr_\pps f \big( \pps + \hh^{+}_\pps, \AA +
J\HH^{+}_\pps \big) \cr & -\e \int_{{\bf T}^l} d\pps \,
\HH^{-}_\pps\cdot\dpr_\AA f \big( \pps + \hh^{+}_\pps, \AA +J
\HH^{+}_\pps \big) + {\L^{-1} \aaa(\e)} \cdot \int_{{\bf T}^l} d\pps \,
\hh^{-}_\pps \cr} \Eq(1.6) $$
where $\aaa$ will be called {\it counterterm}, and its physical
dimensions are $[\oo]$.

It is easy to check that the Schwinger functions
$$ S_n (\pps_1,s_1,\s_1;\ldots;\pps_n,s_n,\s_n) =
{\int P(d\F)\, e^{-V(\F)}\,
\F^{s_1\s_1}_{\pps_1} \ldots \F^{s_n\s_n}_{\pps_n} \over
\int P(d\F)\, e^{-V(\F)} } \Eq(1.7) $$
of the non polynomial formal\footnote{${}^4$}{\nota Because the
$\F$'s are complex and $f$ is a trigonometric polynomial.} action
\equ(1.6) are well defined if the one point Schwinger functions
$$\V h(\pps) \= S_1(\pps,2,+) = {\int P(d\F)\, e^{-V(\F)}\,
\hh^{ +}_{\pps} \over \int P(d\F)\, e^{-V(\F)} } , \qquad \V H(\pps)
\= S_1(\pps,1,+) = J{\int P(d\F)\, e^{-V(\F)}\, \HH^{+}_{\pps} \over \int
P(d\F)\, e^{-V(\F)} } \Eq(1.8) $$
are well defined. The reason is simply that the structure of the free
field and that of the action imply that all the Feynman graphs of the
theory must be either trees or families of disconnected trees.  The
renormalization constant $\V a(\e)$ will be fixed by requiring that the
average of $\V h$ vanishes. As in field theory one could fix $\V a$
equivalently by requiring that the average of $\V h$ has a prefixed
value: it is only important that $\V h$ is well defined when $\L\to\io$
and the value $\V0$ for its average has no special meaning, except that
it is a convenient normalization which, as we shall see, makes use of
the symmetry of the problem inherited by the fact that $f$ has a cosine
Fourier series and this simplifies some considerations.

The case in which $f$ is $\AA$-independent has been studied in [G3],
where it has been shown that in the limit $\L\to\io$ the one point
Schwinger functions are precisely the functions $\V h$ and $\V H$
defined by the KAM theorem, provided the counterterms $\aaa$ are chosen
$\V0$.  In [G3] the $\V a$ does not appear (as it is $\V0$ for symmetry
reasons) so that the analysis is considerably simpler and no cut off
$\L$ is necessary. The necessity of $\aaa\neq\V0$ arises only if $f$ is
$\AA$-dependent (and it is related to the twist condition that becomes
necessary in such a case: note that in [G3] the twist condition was not
required; furthermore, as a consequence, only one field, namely $\V
h_{\pps}$, was used).

In this paper we study the more general case in which the action \equ(1.6)
depends also on $\AA$. If the ultraviolet cut off $\L$ is finite the
perturbative expansion for the Schwinger functions is convergent for
$\e$ suitably small and for any choice of the counterterms. However,
in the limit $\L\to\io$ the series is convergent for a {\it unique}
choice of the counterterm $\aaa(\e)$.  This is what happens
generically in quantum field theory, in which the pertubative series
for Schwinger functions converge only if a unique choice of the
counterterm is made (see for instance the case of $\phi^4$, [G1]).
{\it Moreover the choice of the counterterms which makes the
perturbative series finite in the limit $\L\to\io$ is such that
$\V h$, $\V H$ in \equ(1.8) {\it coincide} with the
corresponding quantities in the KAM theorem.}

\vskip1.truecm
\\{\bf 2. The Schwinger functions expansion.}
\numsec=2\numfor=1
\*
\\The latter statement can proved by writing recursively the one point
Schwinger function to order $n$, $H^{(n)}_{\nn,j}$ and
$h^{(n)}_{\nn,j}$ and comparing it with a similar recursive
construction of the Lindstedt series for the KAM functions $\V H,\V h$.

The exponentials in \equ(1.7) are expanded in powers of $V$ and the
$P$ integrals of the resulting products of fields are evaluated using
the Wick rule leading to the familiar Feynman diagrams: the special
form of $V$ immediately implies that the diagrams have no loops, \ie
they are tree diagrams.

The diagrams will be described later: here it is sufficient to remark
that even without using the diagram representation the evaluation of
the integrals immediately leads to the following recursive relations
between the coefficients of the power series (in $\e$) expansion of
the functions $\V H,\V h$ in \equ(1.8), \ie the one field Schwinger
functions of the theory described by \equ(1.5), \equ(1.6):
$$H^{(k)}_{\nn,j} = S^1_\nn \Big\{ {\sum}^* (-i\nn_0)_j
\sum_{p,q\ge0}\fra1{p!q!}\prod_{s=1}^p
\big(i\nn_0\cdot\V h^{(k_s)}_{\nn_s}\big) \prod_{i=1}^q\big(\V
H^{(k'_i)}_{\nn_i'}\cdot\dpr_{\V A}\big)\, f_{\nn_0}(\AA)
\Big|_{\AA=\AA_0}\Big\}+J a^{(k)}_j\d_{\nn,\V0} \; , \Eq(2.1) $$
and:
$$\eqalign{
&h^{(k)}_{\nn,j} = S^2_\nn \Big\{ {\sum}^*
(-iJ^{-1}\nn_0)_j \sum_{p,q\ge0}\fra1{p!q!}
\;\prod_{s=1}^p \big(i\nn_0\cdot\V h^{(k_s)}_{\nn_s}\big)
\prod_{i=1}^q\big(\V H^{(k'_i)}_{\nn_i'}\cdot\dpr_{\V A}\big)\,
f_{\nn_0}(\AA) \Big|_{\AA=\AA_0}\Big\}
\cr & + \L a^{(k)}_j \d_{\nn,\V0} +
S^1_\nn \Big\{ {\sum}^* \sum_{p,q\ge0}\fra1{p!q!}
\prod_{s=1}^p\big(i\nn_0\cdot\V h^{(k_s)}_{\nn_s}\big) \prod_{i=1}^q
\big(\V H^{(k_{i}')}_{\nn_{i}'}\cdot\dpr_{\AA}\big)
\dpr_{\AA_j}f_{\nn_0}(\AA)\Big|_{\AA=\AA_0}\Big\} \; , \cr}
\Eq(2.2)$$
where the ${\sum}^*$ denotes sum over the integers
$k_{s},k_{i}'\ge1$ and over the integers $\nn_0$, $\nn_{s}$,
$\nn_{i}'$, with
$$ \sum_{s=1}^{p} k_{s} + \sum_{i=1}^{q} k_{i}' = k-1,\qquad
\nn_0 + \sum_{s=1}^{p} \nn_{s} + \sum_{i=1}^{q}
\nn_{i}' = \nn \; . \Eq(2.3)$$
The integer vectors $\V\n_s,\V\n'_i,\V\n_0,\nn$ may be $\V0$.

For $\nn=\V0$, from the above relations we obtain
$$ H_{\V0,j}^{(k)} = \L X_j^{(k)} +J a^{(k)}_j ,\qquad h^{(k)}_{\V0,j}
= J^{-1}\L [ \L X_j^{(k)} + J a^{(k)}_j ] + \L Y_j^{(k)} \; , \Eq(2.4) $$
where $X_j^{(k)}$ e $Y_j^{(k)}$ are read from \equ(2.1) and \equ(2.2)
for $\nn=\V0$. The condition that $\V h_{\V0}^{(k)}=\V0$ determines,
recursively, $a^{(k)}_j$ and implies $\V H_0^{(k)}=-J\V Y^{(k)}$.

The first order calculation yields
$$ \eqalign{
\V H_\nn^{(1)}& = S_\nn^1 (-i\nn)\,f_\nn + J \aaa^{(1)}
\d_{\nn,\V0} \; , \cr
\V h_\nn^{(1)} & = J^{-1}S_\nn^2 (-i\nn)\,f_\nn + S_\nn^1 \aaa^{(1)}
\d_{\nn,\V0} + S_\nn^1  \dpr_\AA f_\nn \; , \cr} \Eq(2.5) $$
and the limit as $\L\to+\io$ is well defined if
$\V a^{(1)}=J^{-1}\V H^{(1)}_{\V0}=-\dpr_{\AA}f_{\V0}(\AA_0)$, and it is
$$ \eqalign{
\V H_\nn^{(1)} & = {(-i\nn)\,f_\nn(\AA_0) \over i\oo\cdot\nn } \; ,
\qquad \V h_\nn^{(1)}  = {(-iJ^{-1}\nn)\,f_\nn(\AA_0) \over
(i\oo\cdot\nn)^2} + {\dpr_\AA f_\nn (\AA_0) \over i\oo\cdot\nn } \; ,
\qquad \nn\neq\V0 \; , \cr
\V H^{(1)}_{\V0} & =-J\dpr_\AA f_\V0 (\AA_0)\; ,\kern1.3cm \V h^{(1)}_{\V0}=
\V0 \; , \kern2cm {\rm if}\ J \V a^{(1)}=H^{(1)}_{\V 0} \; , \cr} \Eq(2.6) $$
with $\V h^{(1)}_{\V0}=\V0$ and the functions $\V H$ and $\V h$
respectively even and odd in $\V\n$, (as in [GM]).

Then, if we want that the expressions in \equ(2.1), \equ(2.2) are well
defined when $\L\to\io$, we proceed inductively by supposing that by
suitably fixing $\V a^{(k)}$ the functions $\V H^{(k)}$ and $\V h^{(k)}$
have a well defined limit as $\L\to+\io$ and become, respectively,
even and odd in $\nn$ when the limit is taken.  We assume this to be
true for $k'\le k-1$: we see that this implies $X^{(k)}_j=0$ in the first
equation, and the choice $a^{(k)}_j=-Y^{(k)}_j$ makes the parity and
finiteness requests to be fulfilled to order $k$.

\vskip1.truecm

\\{\bf 3. The Lindsted series.}
\numsec=3\numfor=1
\*
\0The classical construction of the formal series representation for
the functions $\V H,\V h$ in \equ(1.3) defining parametrically the
KAM torus starts from the Hamilton equations of motion for \equ(1.1).
One imposes that by replacing $\pps$ with $\pps+\oo_0 t$ in \equ(1.3)
one gets an exact solution to the equations of motion. The following
equations are obtained:
$$ \eqalign{
\oo_0\cdot\V\dpr_\pps\,\V H(\pps)=&-\e\,\dpr_\pps f
\left(\pps+\V h(\pps),\AA_0+\V H(\pps)\right) \; , \cr
\oo_0\cdot\V\dpr_\pps\,\V
h(\pps)=&J^{-1}\V H(\pps)+ \e\,\dpr_\AA f\left(\pps+\V h(\pps),\AA_0+\V
H(\pps)\right) \; . \cr}\Eq(3.1)$$

To make easier the comparison with the euclidean field theory of \S2 we
can introduce a cut off parameter $\L$ and consider the regularized
equations
$$\eqalign{
(\L^{-1}+\oo_0\cdot\dpr_\pps)\,\V H(\pps)=&-\e\,\dpr_\pps f\left(\pps+\V
h(\pps),\AA_0+\V H(\pps)\right) \; , \cr
(\L^{-1}+\oo_0\cdot\V\dpr_\pps)\,\V h(\pps)=&J^{-1}\V H(\pps)+\e\,
\dpr_\AA f \left(\pps+\V h(\pps),\AA_0+\V H(\pps)\right) \; . \cr}\Eq(3.2)$$

We can solve \equ(3.2) by a perturbation expansion, by writing $\V
H=\sum_{k=1}^\io \e^k\V H^{(k)}$ and $\V h=\sum_{k=1}^\io \e^k\V
h^{(k)}$.  If one requires $\V h^{(k)}_{\V0}=\V0$ then it follows
immediately that the recursive construction of $\V H^{(k)},\V h^{(k)}$
is possible and in fact it clearly coincides with
\equ(2.1)$\div$\equ(2.6). The existence of such formal series is known
(if $\L=+\io$) as the {\it Lindstedt theorem}: and it goes back to
Poincar\'e who extended to all orders the original proofs of Lindstedt
and Newcomb.

The convergence radius of the Lindstedt series (hence of the euclidean
field theory of \S2) is uniform in $\L$. For $\L=+\io$ this is the KAM
theorem; a proof based on bounds on the coefficients $\V H^{(k)},\V
h^{(k)}$ is due to Eliasson, [E]. It was recently ``simplified" in various
papers [G2], [GG], [GM], see also [CF] for a very similar approach. The
proof in [G2], [GM] can be easily extended to cover the case
$\L<+\io$. Hence the theory is {\it uniform} in the ultraviolet cut
off $\L$ (of course the convergence at fixed $\L<\io$ is quite
trivial; the uniformity as $\L\to\io$, on the other hand, is
equivalent to KAM).

\vskip1.truecm

\\{\bf 4. The renormalization group and resonance resummation.}
\numsec=4\numfor=1 \*

\0The KAM theory, thus, permits us to give a meaning to the non
regularized field theory with action \equ(1.6), a somewhat
surprising fact.  Therefore it is interesting to investigate in more
detail the structure of the perturbation theory.

As already pointed out the model is, from the point of view of field
theory, somewhat deceiving as its Feynman diagrams have no loops.
Nevertheless the model is clearly non trivial and it requires a
delicate analysis of a family of cancellations that make the
ultraviolet stability possible at all.

With the choice of the counterterm $\aaa(\e)$ las in \S 2
the Feynman rules for the model can be summarized as follows.
Consider $k$ oriented lines, labeled from $1$ to $k$: the final extreme
$v'$ of the lines will be called the {\it root} and the other extreme
$v$ will be a {\it vertex}. The lines, denoted $v'\ot v$ are arranged
on a plane by attaching in all possible ways the vertices of some
segments to the roots of others, to form a connected tree.

In this way only one root $r$ remains unmatched and it will be
called the root of the graph whose lines will be called {\it
branches} and whose vertices other than the root will be called {\it
nodes}.

Each node $v$ is given a {\it mode} label $\nn_v$ which is one of the
Fourier mode $\nn$ such that $f_\nn\ne0$ (see \equ(1.1)). We define the
{\it momentum} flowing on the branch going from $v$ to $v'$ as
$\nn(v)=\sum_{w\le v}\nn_w$. Furthermore each branch is regarded as
composed by two halves each carrying a label $H$ or $h$ (so there are
four possibilities per branch).

Trees that can be superposed modulo the action of the group of
transformations generated by the permutation of the branches emerging
from a node will be identified.

To each tree we associate a {\it value} obtained by assigning to a
branch $v'\ot v$ the following quantities, if $\nn(v)\neq\V0$,
\halign{#\hfill\kern0.5cm
&\hfill#\hfill&\kern2cm#\cr
& & \cr
{a factor}&{ $\dt \fra{-i\nn_{v'}\cdot
iJ^{-1}\nn_{v}}{(i\oo_0\cdot\nn(v)+\L^{-1})^2}$}&  {$h\ot h$}\cr
& & \cr
{an operator}&{
$\dt \fra{i\nn_{v'}\cdot\dpr_{\AA_{v}}}
{i\oo_0\cdot\nn(v)+\L^{-1}}$}&  {$h \ot H$}\cr
& & \cr
{an operator}&{ $\dt \fra{-\dpr_{\AA_{v'}}
\cdot i\nn_v}{i\oo_0\cdot\nn(v)+
\L^{-1}}$} & {$H \ot h$}\cr
& & \cr
{just}&$\dt 0$&{$H \ot H$}\cr
& & \cr}
\\for all the branches distinct from the one containing the root: here
the symbol to the right distiguishes the four type of labels that can be
on the line $v'\ot v$ (the arrow tells which is the right label and
which is the left one). To the root branch we associate, instead, the
following quantities, if $\nn(v)\neq\V0$,
\halign{#\hfill\kern0.5cm&\hfill#\hfill&\kern2cm#\hfill\cr
& & \cr
{a vector}&{
$\dt\fra{-iJ^{-1}\nn_v}{(i\oo_0\cdot\nn(v)+\L^{-1})^2}$} & {$h \ot h$}\cr
& & \cr
{an operator}&{ $\dt\fra{\dpr_{\AA_v}}{i\oo_0\cdot\nn(v)+\L^{-1}}$}& {$h\ot
H$}\cr & & \cr
{a vector}&{ $\dt\fra{-i\nn_v}{i\oo_0\cdot\nn(v)+\L^{-1}}$}
&{$ H\ot h$}\cr & & \cr
{just}&$\dt0$& {$H \ot H$}\cr& & \cr}
To each branch with $\nn(v)=0$ which is not the root branch we associate
a factor $-J\dpr_{\AA_{v}}\cdot \dpr_{\AA_{v'}}$, if ${H\ot h}$, and $0$
otherwise, while to the root branch we associate a factor
$-J\dpr_{\AA_{v}}$, if $H\ot h$, and $0$ otherwise.

We multiply all the above operators (the factors are regarded as
multiplication operators) and apply the resulting operator to the
function $\prod_v f_{\nn_v}(\AA_v)$, evaluating the result at the point
$\AA_v\=\AA_0$.  This defines the Feynman rules: the $\V H^{(k)}_\nn$
and $\V h^{(k)}_\nn$ are given by $k!^{-1}$ times the sum of all the
values of all the $k$ branches trees with total momentum $\nn$. In the
limit $\L\to\io$, the above expressions are all well defined: this is
easily checked.  The expansion was developed in [G2], [GM] and it
coincides essentially with the one used in [E] (and [CF]).

Note that, in [GM], each time a line $\l$ carries a vanishing momentum,
all the subtrees of fixed order $k_1$ having $\l$ as first branch are
summed together to give, by construction, the value of the counterterm
$\aaa^{(k_1)}$. Such a contribution is called {\it fruit} in [GM], and
a line of a fruitful tree can have vanishing momentum only if it comes
out from a fruit. Obviously the two ways to arrange the sums over the
trees are equivalent, and give the same result, {\it once the sums are
extended to all the possible trees}.


The scaling properties of the propagators (when $\L=+\io$) suggest
decomposing them into components relative to various {\it scales}.

Let $\chi_1,\chi$ be two smooth functions such that:

\0(1) $\chi_1(x)\=0$ if $|x|<1$ and $\chi_1(x)\=1$ for $|x|\ge1$.

\0(2) $\chi(x)\=0$ for $|x|<\fra12$ or for $|x|\ge1$, and $1$ otherwise.

\0(3) $1\=\chi_1(x)+\sum_{n=-\io}^0\chi(2^n x)$

Then we can write:
$$S^a_\nn\=\fra1{(i\oo_0\cdot\nn)^a}=
\fra{\chi_1(\oo_0\cdot\nn)}{(i\oo_0\cdot\nn)^a}+ \sum_{n=-\io}^0
\fra{\chi(2^{-n}\oo_0\cdot\nn)}{(i\oo_0\cdot\nn)^a}
\; , \qquad a=1,2, \Eq(4.1)$$
and correspondingly we can break each Feynamn graph into a sum of
many terms by developing the sums in \equ(4.1). This can be simply
represented by assigning to each branch $\l$ an extra label $n_\l$
and multiplying the factor associated to such a line times
$\chi(2^{-n_\l}\oo_0\cdot\nn)$:
the value of $\V H^{(k)}_\nn,\V h^{(k)}_\nn$ will be the sum
over all possible new graphs which once deprived of the new scale
labels would become ``old" graphs contributing to
$\V H^{(k)}_\nn,\V h^{(k)}_\nn$  respectively.

The branches of the new graphs are naturally collected into connected
{\it clusters} ``of fixed scale": a cluster of scale $n$
($n=1,0,-2,\ldots$) consists in a maximal connected set of branches
with scale label $\ge n$, containing at least one line of scale $n$.
By definition each cluster is again a tree graph. The lines which are
not contained in a cluster, but have an extreme inside the clusters
will be called the external lines of the cluster: if the extreme
inside the resonance is the root, they will be {\it incoming},
while if the extreme is the node they will be {\it outgoing}.
There can be at most one outgoing line per cluster.

The clusters are, by definition, hierarchically ordered and therefore
they form a tree with respect to the partial ordering generated by
the inclusion relation between clusters.

Examining the convergence of the perturbation series it becomes clear
that if one considers the sum of the contributions to $\V H^{(k)},\V
h^{(k)}$ by all the graphs that {\it do not contain clusters with
just one incoming and one outgoing branch which, furthermore, have
the same momentum $\nn$}, then the series so generated converge for
$\e$ small, [E], [FT].

Therefore the clusters of the latter type (with one incoming and one
outgoing equal momentum branches) are called {\it resonances} and the
KAM theory can be interpreted as an analysis of the reason why the
resonances do not destroy the analyticity in $\e$ at $\e$ small, \ie
of the cancellations that make the resonances give a
contribution {\it much smaller} than one could fear.

One can imagine to consider a graph and replace each resonance
together its external lines with a new simple line, which will
be called {\it dressed line}. We collect togheter all the graphs
which become identical after such an operation.

We consider here for simplicity only the case in which $f$ is $\AA$
independent; the discussion of the more general case, $f=f(\aa,\AA)$,
can be carried out in the same way and it is only notationally more
involved, so that, for simplicity's sake, we relegate it into Appendix
A2.  If we multiply each graph value by the appropriate power of $\e$
(equal to the number of branches of the graph) we see that the values of
$\V H$ and $\V h$ can be computed by considering all the graphs without
resonances and by adding resonant clusters to each of their lines. This
simply means that a line factor of scale $n$ has to be modified as:

$$\chi(2^{-n}\oo_0\cdot\nn(v))\fra{
(-i\nn_{v'}\cdot iJ^{-1}\nn_v)}{(i\oo_0\cdot\nn(v))^2}\ \to\
\fra{\chi(2^{-n}\oo_0\cdot\nn(v))}{(i\oo_0\cdot\nn(v))^2}
(-i\nn_{v'}\cdot
[(1-\s_{n,\e}(\oo_0\cdot\nn(v))]^{-1}iJ^{-1}\nn_v)\Eq(4.2)$$
where $\sigma_{n,\e}(\oo_0\cdot\nn)$ is a suitable function representing
the sum of all the possible insertions of a resonant cluster on the line
$v'\ot v$. The function $\s_{n,\e}(\oo_0\cdot\nn)\equiv\s_{n,\e}(2^{n}
x)$ does not vanish only for $x$ in the interval $[\fra12,1]$.

The following result is an immediate consequence of the
results in [G2], [GM2].
\*

\0{\bf Theorem.} {\it The matrix $\s_{n,\e}(2^{n}x)$ is analytic in $\e$
for $\e$ small, independently on $n$ and there is a constant $R$ such
that $||\s_{n,\e}(2^{n}x)||< R |\e|$.\hfill\break Furthermore the limit:
$$\lim_{n\to-\io} \s_{n,\e}(2^n x)=\s_\e\Eq(4.3)$$
exists and is a $x$--independent function of $\e$, analytic for
$\e$ small enough and divisible by $\e$.}
\*

The second part of the above theorem is discussed in Appendix A1.
The first part is proven in [GM2] in a version in
which the $\chi$ functions are not characteristic functions as above,
but are smoothed versions at least two times differentiable. However
one can easily take them to be as above: this implies that when they are
differentiated their derivatives have to be interpreted as combinations
of delta functions. But one checks that most of of such terms cancel
with each other with some obvious exceptions which can be easily
bounded. The possibility of using characteristic functions in the
decomposition \equ(4.1) can also be seen from [G2], where the
decomposition is done as above.
The constant matrix $\s_\e$ will be called the {\it resonance form
factor}.

It is natural to consider the two parameters series $\V
H^*(\pps,\e,\s), \V h^*(\pps,\e,\s)$ obtained from the resonance
resummed series by replacing $\s_{n,\e}$ by a {\it new, independent}
parameter $\s$. Then the above theorem and the results of
[G2],[GG],[GM] imply that the functions $\V H^*,\V h^*$ are analytic
both in $\e$ and $\s$ near the origin.

In fact it is clear that the functions $\V H^*,\V h^*$ depend only on
the variables $\h=\e (1-\s_\e)^{-1}$. Thus the possibility arises that a
singularity for $\V H^*,\V h^*$ is reached at a value $\e_c$ of $\e$
where $\s_\e$ is {\it still finite}.  It seems natural, to us, to think
that the singularities of the functions $\V h, \V H$ as $\e\to\e_c$ are
the same as those of $\V h^*,\V H^*$.  If so the breakdown of the torus
can be studied by using for it a much simpler perturbation
representation, \ie a representation in which no resonance appears in
the graphs representing the $\V H^*,\V h^*$.
\vskip0.5cm

\0{\bf 5. Heuristic discussion of a possible universality mechanism for
the brakdown of the tori.}
\numsec=5\numfor=1\*

\\The scalar quantity $\s_\e$ plays the role of a stability indicator and
it would be nice to see some independent physical interpretation of
it. A numerical study of the function $\s_\e$ appears highly desirable,
as well as that of the functions $\V H^*, \V h^*$.

The possibility that the singularities of $\V H^*, \V h^*$, as functions
both of $\e$ and $\pps$, have a {\it universal nature} becomes clear
because the behaviour of the large order coefficients of $\V H^*,\V h^*$,
as series in $\e$, is likely to be very mildly dependent on the actual
values of the Fourier components $f_\nn$. This can be seen to happen
when only the contributions to the coefficients arising from simple
classes of trees are taken into account.

The simplest class of graphs which does not give a trivial contribution,
\ie contribution which is an entire function of $\e$, to the invariant
tori is given by the set of trees of the form ({\it linear chains}):

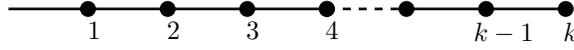
\figini{xyz}
\8<
\8<gsave>
\8</punto { gsave >
\8<3 0 360 newpath arc fill stroke grestore} def>
\8</puntino { gsave >
\8<2 0 360 newpath arc fill stroke grestore} def >
\8<0 20 moveto>
\8<30 20 lineto 30 20 punto>
\8<60 20 lineto 60 20 punto>
\8<90 20 lineto 90 20 punto>
\8<120 20 lineto 120 20 punto>
\8<stroke [3] 0 setdash>
\8<120 20 moveto>
\8<150 20 lineto stroke [] 0 setdash>
\8<120 20 punto 150 20 punto>
\8<150 20 moveto 180 20 lineto 180 20 punto>
\8<210 20 lineto 210 20 punto>
\8<stroke>
\8<grestore>
\figfin

\eqfig{210pt}{40pt}{
\ins{30pt}{15pt}{$1$}\ins{60pt}{15pt}{$2$}
\ins{90pt}{15pt}{$3$}\ins{120pt}{15pt}{$4$}\ins{176pt}{15pt}{$k-1$}
\ins{210pt}{15pt}{$k$}}{xyz}{}

We consider the contribution to $\V h^*(\pps,\e,\s)$ due to the above
trees. For simplicity we fix $l=2$, $\oo=(r,1)$ with $r={\sqrt{5}-1\over
2}=$ {\it golden section} and the perturbation as an even function of
$\V\a$ only as $f(\V\a)=a \cos\a_1+ b\cos (\a_1-\a_2)$ (``Escande Doveil
pendulum'').

Let us call ``resonant line" the line ortogonal to $\oo$, \ie parallel to
$(1,-r)$. Let $(p_n,q_n)$ be the convergents for continued fraction for
$r$ (\ie $p_1=1, p_2=1, p_3=2,...=$ {\it Fibonacci sequence}, and $q_1=1,
q_2=2,q_3=3,\ldots$ with $q_n=p_{n+1}$ and $p_{n+1}=p_n+p_{n-1}$, and we
set $p_0=q_{-1}=0$ and $p_{-1}=q_0=1$).

Any integer $s\ge 1$ can be written:

$$s=q_n+\sigma_{n-2}
q_{n-2}+\ldots+\sigma_1 q_1\Eq(5.1)$$

\0if $q_n\le s< q_{n+1}$ and
$\sigma_1,\sigma_2,\ldots,\sigma_{n-2}=0,1$,
with the constraint $\s_j\s_{j+1}=0$, $j=1,\ldots,n-3$.
Let $\Lambda_{q_n}$ be the family of self avoiding walks on the integer
lattice ${\bf Z}^2$ starting at $(0,0)$, ending at $(q_{n},-p_n)$ and
contained in the strip $0<x\le q_n$, except for the left extreme
points. Then a self avoiding walk joining $(0,0)$ to $(s,s')$ with $s$
given by \equ(5.1) and $s'=p_n+\sigma_{n-2}p_{n-2}+\ldots+\sigma_1 p_1$
can be obtained by simply joining a path in $\Lambda_{q_n}$, one in
$\Lambda_{q_{n-2}}$ if $\sigma_{n-2}=1,\ldots,$ one in $\Lambda_1$ if
$\sigma_1=1$. The latter self-avoiding walks will define the class
$\Lambda_s$ of walks. It is clear by the construction that the above
class $\Lambda_s$ of self avoiding walks contains many of the ones which
have the largest products $\prod_j {1\over (i\oo\cdot\nn(j))^2}$ of
small divisors. Therefore we define:
$$\V Z(\Lambda_{q_n})=\sum_{\rm paths\ in\ \Lambda_{q_n}} (-i
\h J^{-1}\nn_{{1}}) { f_{\vec
\n_{{1}}}\over(i\oo\cdot\vec\n(1))^2}
\prod_{j=2}^{k}{f_{\vec\n_{v_j}}(\vec \n_{j-1}\cdot \h
J^{-1}\vec\n_{{j}})\over
(\oo\cdot\vec\n(j))^2}e^{i(q_n\psi_1-p_n\psi_2)}\Eq(5.2)$$
(with $\nn(1)=(q_n,-p_n)$ which can be always realized with
the vectors $\nn_1=(1,0)$ and $\nn_2=(1,-1)$).

We expect:
$$\V Z(\Lambda_{q_n})=\V\z\
{C(\h,f)^{q_n}\over q_n^\d} \, e^{i(q_n\psi_1-p_n\psi_2)}(1+O(
q_n^{-1}))\equiv\V\z Z(\L_{q_n})
= Z_n\,e^{i(q_n\psi_1-p_n\psi_2)} \; , \Eq(5.3)$$
where $\V\z$ is a suitable unit vector, $C(\h,f)$ is a suitable function
of $\h f$ and $\d$ is a critical exponent characteristic of the golden
section. Then the contribution to $\V h^*$ due to the above classes of
trees and paths can be computed approximately, by noting that, if
$\e_n=(r p_n- q_n)$ and $Z(\L_s)$ is defined as in \equ(5.2), \equ(5.3),
with the sum being over the paths in $\L_s$ and $\nn(1)=(s,-s')$,
$$\eqalign{ Z(\L_s)\simeq& Z(\Lambda_{q_n})
Z(\Lambda_{q_{n-2}})^{\sigma_{n-2}}\ldots Z(\Lambda_{q_1})^{\sigma_1} \;
, \qquad s<q_{n+1} \; , \cr Z(\Lambda_{q_{n+1}})\simeq&
Z(\Lambda_{q_n})Z(\Lambda_{q_{n-1}})
\Big({\e_{n-1}\over\e_{n+1}}\Big)^2 \; , \qquad
s=q_{n+1} \; , \cr}\Eq(5.4)$$
where we can define, for consistency, $Z(\L_{q_0})=\e_0^{-2}=r^{-2}$ and
$Z(\L_{q_{-1}})=(\e_0/\e_{-1})^2=r^2$.  This means that the contribution
to $\V h^*$ can be written approximately, if $r_n={p_n\over q_n}$:
$$\eqalign{&\vec\z\,\sum_{n=1}^\io
\sum_{\sigma_1,\ldots,\sigma_{n-2}=0,1}
\,Z_n\,Z_1^{\sigma_1}T_{\s_1\s_2}\, Z_2^{\sigma_2}
T_{\sigma_2\sigma_3}\ldots T_{\sigma_{n-3}\sigma_{n-2}}
Z_{n-2}^{\sigma_{n-2}} \, e^{i(s\psi_1-s'\psi_2)}\cr
&\simeq\vec\z\,\sum_{n=1}^\io Z_n e^{i(q_n\psi_1-p_n\psi_2)}
{\rm\,Tr\,}[\Th_1\Th_2\ldots\Th_{n-2}] \; , \cr}\Eq(5.5)$$
where $T_{\s\s'}$ is the compatibility matrix defined to be $T_{11}=0$,
$T_{00}=T_{01}=T_{10}=1$, and $\Th_j$, $j=2,\ldots, n-2$, are defined as
$(\Th_j)_{\s\s'}=T_{\s\s'} Z_j^{\s'}$ and $(\Th_1)_{\s\s'}=Z^{\s'}_1$.

If $T_{\s\s'}$ were $\=1$ the trace would be simply $\prod_j
(1+C(\h,f)^{q_j}q^{-\d}_j)$; so that, in the above approximation the
series will become singular when $|C(\h,f)|=1$ and in that case
${\rm\,Tr\,}[\Th_1\Th_2\ldots\Th_{n-2}]$ can probably be replaced by a
constant, as far as the determination of the singularity in $\V\psi$ is
concerned (and perhaps in $\h$ or $\e$ as well). Hence we find the
following representation of the contribution to $\V h^*$ that we are
considering:
$$\vec\z\sum_{n=1}^\io { [C(\h,f)e^{i(\psi_1-r_n\psi_2)}]^{q_n}\over
q_n^\d}=\vec\z \sum_{n=1}^\io {[C(\h,f) e^{i(\psi_1-r\psi_2)}]^{q_n}
\over q_n^\d}e^{-i\psi_2 O({1\over q_n})} \; . \Eq(5.6)$$

We expect that the singularities of $\V H^*,\V h^*$, as $\e$ grows, are
the same as those of $\V H,\V h$, and furthermore we expect the above
considered contributions to the functions $\V H^*, \V h^*$ to be the
{\it most singular}. Hence we interpret \equ(5.6) as saying that we
should expect $\V h,\V H$ to be, at the breakdown of the invariant torus
which corresponds to $|C(\h,f)|=1$, singular as functions of
$\psi_1,\psi_2$ and of $\h$ (hence of $\e$).

Furthermore the set $|C(\h,f)|=1$ is in the $\h$-plane a natural
boundary for the functions $\V h^*,\V H^*$ as functions of $\h$ and, if
$\h={\e\over 1-\sigma_\e}$ is smooth in $\e$, or at least
Lipshitz continuous, as mentioned above, when $\e\to\e_c^-$,
$C(\h,f)\simeq(1-\g(\e_c-\e))$ so that the singularity of $\V
h(\psi_1,0)$ or $\V H(\psi_1,0)$ in $\e,\psi_1$ is described by the
singularity of a single function $\x(z)$, or $\x'(z)$, of the single
variable $z=e^{-\g(\e_c-\e)} e^{i\psi}$:
$$\x(z)=\sum_{k=1}^\io \fra{z^{q_k}}{q_k^\d},\qquad{\rm or}\quad
\x'(z)=\sum_{k=1}^\io \fra{z^{q_k}}{q_k^{\d+1}}\Eq(5.7)$$
which would mean that the {\it critical torus} has a Lipshitz continuous
regularity with {\it any} exponent $\d'<\d$ in the $\psi_1$--variable
and $\d$ in the $\e-\e_c$ variable, [K].

For instance if we fix $\psi_2$, \eg as $\psi_2=0$ (which can be
regarded as a special Poincar\'e section of the invariant torus), then
$\V h^*$ is a $C^{\d'}$ function and $\V H^*$, which is obtained from $\V
h$ by applying the operator $\V\o\cdot\dpr_{\pps}$, is a $C^{1+\d}$
function, [K].  Note that the structure of the operator
$\V\o\cdot\dpr_{\pps}$ is such that when it is applied to $\V h$ as in
\equ(5.6) it generates a {\it smoother function}.  Therefore, based on
the hypothesis that the singularity of $\V H,\V h$ and of
$\V H^*,\V h^*$ are the same, see \S4, and on the above heuristic
discussion, the following conjecture emerges.  \*

\0{\it Conjecture.  Consider the conjugacy to a pure rotation of the
motion generated by the Poincar\'e map on a circle on the critical
torus.  There is $\d>0$ such that
it is described by two functions $\V h,\V H$ and written as
$\V\a=(\psi,0)+(h_1(\psi),h_2(\psi))$ and $\V A=(H_1(\psi),H_2(\psi))$
with $\V h$ H\"older continuous with exponent $\d'<\d$ and $\V H$ of
class $C^{1+\d'}$.  Furthermore the above conjugacy has a H\"older
continuous regularity $\d'<\d$ in the $\e-\e_c$ variable.} \*

The mechanism for universality in the breakdown of the invariant tori
that we propose above is, in our opinion, a refined version of an
important idea in [PV]: except that we have {\it not} made here the
simplifying assumption of absence of resonances (\ie we {\it allow} for
non zero Fourier components of opposite wave label $\pm\nn$, and find
resummations that in some sense eliminate them).

If one accepts that the above pendulum system has the same critical
exponents for the golden mean torus in the standard map then it follows
that $\d=0.7120834$ by the scaling argument on p.207 of
[Ma].\footnote{${}^1$}{\nota Private communication of MacKay.} The
regularity of the two conjugators is in fact in that case not smoother
than $C^\d$ for the analogue of $\V h$ and of $C^{1.9568}$ for the
analogue of $\V H$: hence the above conjecture is in agreement with the
data and gives some independent reasons for the difference of about $1$
between the regularity of $\V h$ and that of $\V H$. Unfortunately an
exact computation of the regularity of $\V H$ does not seem to have bee
attempted yet.\footnote{${}^2$}{\nota Private communication of MacKay.}
\vskip1.truecm

\\
{\bf Appendix A1.} {\it The stability constant $\s_\e$.}
\*

\\We fix $n$ and we consider the contribution to $\s^{(k)}_{n,\e}(2^nx)$
arising from a $k$-th order term corresponding to a given Feynman graph:
it will be given by the sum of products of factors whose dependence on
the variable $2^nx$ is through terms of the form:

$$ (\oo_0\cdot(\nn^0_\l+2^nx))^{-1} \; , $$

\0where $\nn^0_\l$ is the momentum of the branch $\l$ inside the
resonance, \ie the sum of all the modes of the vertices preceding $\l$
contained in the resonance.  Then $|\nn^0_\l|\leq kN$ and by the
diophantine property $|\oo_0\cdot\nn^0_\l|> [C_0 (kN)]^{-\t}$ so that
$n_\l>\tilde n=-\t\log(kN)-\log C_0$, for all $\l$ inside the
resonance. Then, if $k$ is fixed and $n\to-\io$, the quantity
$|\oo_0\cdot\nn^0_\l|$ remains bounded from below because
$|\oo_0\cdot\nn_0|\ge 2^{\tilde n}$ while $2^nx\to 0$ and the
$x$--dependence is only via quantities like $(\oo_0\cdot\nn_\l+ 2^n
\s x)$, $\s=0,1$.  Therefore the dependence on $x$ disappears, and we have:
$$\lim_{n\to\io}\sigma_{n,\e}^{(k)}(2^n x)=\sigma_{\e}^{(k)} \; . $$
On the other hand, as $\sigma_{n,\e}^{(k)}(2^n x)$ is a power series in
$\e$ uniformly convergent, see [GM], and we can pass to the limit under
the sign of series and the theorem is proven.

\vskip1.truecm

\\{\bf Appendix A2.} {\it Resonance form factors for an action dependent
interaction}
\*

\\In general the interaction potential depends also on the action
variables. This yields that all the line factors introduced in \S 4
are possible, so that to the dressed lines we associate the
following quantities
\halign{#\hfill\kern0.5cm
&\hfill#\hfill&\kern2cm#\cr
& & \cr
{a factor}&{ $\dt \chi(2^{-n}\oo_0\cdot\nn(v))\,
\fra{-i\nn_{v'}\cdot
[1-\s^s_{n,\e}(\oo_0\cdot\nn(v))]^{-1}
iJ^{-1}\nn_{v}}{(i\oo_0\cdot\nn(v)+\L^{-1})^2}$}& {$h\ot h$}\cr & & \cr
{an operator}&{ $\dt \chi(2^{-n}\oo_0\cdot\nn(v))\,
\fra{i\nn_{v'}\cdot
[1-\s^s_{n,\e}(\oo_0\cdot\nn(v))]^{-1}
\dpr_{\AA_{v}}} {i\oo_0\cdot\nn(v)+\L^{-1}}$}&  {$h \ot H$}\cr
& & \cr {an operator}&{ $\dt \chi(2^{-n}\oo_0\cdot\nn(v))\,
\fra{-\dpr_{\AA_{v'}}\cdot
[1-\s^s_{n,\e}(\oo_0\cdot\nn(v))]^{-1}
i\nn_{v'}}{i\oo_0\cdot\nn(v)+
\L^{-1}}$} & {$H \ot h$}\cr
& & \cr{just}&$\dt 0$&{$H \ot H$}\cr & & \cr}
\\where $n$ is the scale label of the line, and
$\s^s_{n,\e}(\oo_0\cdot\nn)$, $s=1,\ldots,4$, will have a different form
depending on the labels ($H$ or $h$) attached to the half branches
contributing to form, respectively, the outgoing and the incoming
external lines of the resonant clusters whose values add to
$\s^s_{n,\e}(\oo_0\cdot\nn)$. The analysis in [GM] applies to all kinds
of resonance, so that a result analogous to the theorem of \S4 holds
for all the functions $\s^s_{n,\e}(\oo_0\cdot\nn)$, and four resonance
form factors can be shown to be well defined and depending only on $\e$:
the proof can be carried out exactly in the same way.

\vskip1.truecm

\0{\bf Acknowlegments: \it We are indebted to G. Parisi for an early
suggestion discussed in [G3] and to R. MacKay for explaining us the
results on the standard map.  This work is part of the research program
of the European Network on: ``Stability and Universality in Classical
Mechanics",} \# ERBCHRXCT940460.

\vskip1.truecm

{\bf References}
\*
\0{[CF] } Chierchia, L., Falcolini, C.: {\it A direct proof of a theorem
by Kolmogorov in hamiltonian systems}, Annali della Scuola Normale
Superiore di Pisa, {\bf 21}, 541--593, 1994.

\0{[E] } Eliasson, L.H.: {\it Absolutely convergent series expansions
for quasi periodic motions}, report 2-88, Dept. of Mathematics,
University of Stockholm, 1988.

\0{[FT] } Feldman, J., Trubowitz, E.: {\it Renormalization
in classical mechanics and many body quantum field theory}, Journal
d'Analyse Math\'ematique, {\bf 58}, 213-247, 1992.

\0{[G1] } Gallavotti, G.: {\it Renormalization Theory and Ultraviolet
Stability for scalar Fields via Renormalization Group Methods},
Reviews of Modern Physics, {\bf 57}, 471--562, 1985.

\0{[G2] } Gallavotti, G.: {\it Twistless KAM Tori}, Communications in
Mathematical Physics, {\bf 164}, 145--156, 1994.

\0[G3] Gallavotti, G.: {\it Perturbation Theory}, in ``Mathematical
physics towards the XXI century", p.  275--294, ed.  R.  Sen, A.
Gersten, Ben Gurion University Press, Ber Sheva, 1994.

\0{[GG] } Gallavotti, G., Gentile, G.: {\it Majorant series for
the KAM theorem}, in {$\tt mp\_arc@math.utexas.edu$}, \#93-229, to appear
in Ergodic Theory and Dynamical Systems.

\0{[GM] } Gentile, G., Mastropietro, V.: {\it Tree expansion and multiscale
decomposition for KAM tori}, Roma 2, CARR--preprint 8/9, 1994.

\0{[K]} Katznelson, Y.: {\it An introduction to harmonic analysis},
Dover, 1976.

\0{[Ma]} MacKay R.: {\it Renormalization in area preserving maps}, World
Scientific, London, 1993.

\ciao